\documentclass[twocol]{ametsocV6.1}

\usepackage{caption, subcaption, float}
\usepackage{subeqnarray}
\usepackage{gensymb}
\usepackage{multirow}
\allowdisplaybreaks
\usepackage{color}
\usepackage{booktabs}

\newcommand{\R}{\mathbb{R}}

\newcommand{\derp}[2]{\frac{\partial #1}{\partial #2}}
\newcommand{\der}[2]{\frac{d #1}{d #2}}

\newcommand{\E}{\mathbb{E}}
\newcommand{\Prb}{\mathbb{P}}

\newcommand{\bse}{\begin{subeqnarray}}
\newcommand{\ese}{\end{subeqnarray}}

\DeclareMathOperator*{\argmin}{arg\,min}
\DeclareMathOperator\erf{erf}

\title{Optimal Transition Paths for AMOC Collapse and Recovery in a Stochastic
Box Model}

\authors{Jelle Soons,\aff{a}\correspondingauthor{Jelle Soons j.soons@uu.nl} 
 Tobias Grafke,  \aff{b} 
and Henk A. Dijkstra\aff{a} 
}

\affiliation{\aff{a}{Institute for Marine and Atmospheric Research, Utrecht University, Princetonplein 5, 3584 CC Utrecht, The Netherlands}\\
\aff{b}{Mathematics Institute, University of Warwick, Coventry CV4 7AL, United Kingdom}
}

\abstract{There is strong evidence that the present-day Atlantic Meridional Overturning Circulation (AMOC) is in a bi-stable regime  and hence it is important to determine  probabilities  and  pathways for noise-induced transitions between its equilibrium states.  Here, using Large Deviation Theory (LDT),  the most  probable transition pathways for the noise-induced collapse and recovery of the AMOC are computed in a  stochastic box model of the World Ocean. This allows us to determine the physical mechanisms of noise-induced AMOC transitions. We show  that the  most likely path of an AMOC  collapse starts paradoxically with a strengthening of the AMOC followed by an  immediate drop within a couple of years due to a short but relatively strong freshwater pulse. The recovery on  the other hand is a slow process, where the North Atlantic needs to be gradually salinified over a  course of 20 years. The proposed method provides several benefits, including an estimate of probability ratios of collapse between various freshwater noise scenarios, showing that the AMOC is most vulnerable to freshwater forcing into the Atlantic thermocline region. Moreover, a comparison with a quasi-equilibrium approach reveals the contrasts in behavior of a bifurcation-induced and a noise-induced collapse of the AMOC.}
\begin{document}

\maketitle

%
%
%
%
%

%
\section{Introduction}
The Atlantic Meridional Overturning Circulation (AMOC) is of paramount importance to Earth's climate \protect\citep{srokosz2023atlantic}. It consists of a 
northward transport of warm upper ocean water  and a deeper southward flow of cool water. This northward heat transport is a major reason for the relatively mild climate in Western Europe. An important process behind this circulation is thought to be 
the water mass transformation in the subpolar North Atlantic Ocean. Here,  the relatively warm and salty sea water transported 
from low latitudes is cooled by the atmosphere, and becomes the cold and salty North Atlantic Deep Water (NADW), which 
sinks  and returns southward \protect\citep{frajka2019atlantic}.  

\protect\citet{stommel1961thermohaline} realized that the AMOC is affected by two competing feedback effects due to the coupling of 
the circulation with the density field.  When the AMOC strengthens, more heat will be transported northwards which will 
reduce deep water formation, providing a negative feedback via this thermal circulation. On the other hand, this strengthening also increases
northward salt transport which increases deep water formation and hence is a positive feedback, the salt-advection 
feedback \protect\citep{Marotzke2000}.  These feedbacks, and that the atmosphere damps ocean 
temperature anomalies much more strongly than salinity anomalies, create a multiple equilibrium regime for the 
AMOC. In the box model used in \protect\citet{stommel1961thermohaline}, two stable  equilibria exists under the same 
buoyancy forcing conditions. Such multiple equilibrium regimes have been found  in a number of other box models of the AMOC \protect\citep{rooth1982hydrology, 
rahmstorf1996freshwater, lucarini2005thermohaline, cimatoribus2014meridional} for which the steady states can be computed
when parameters are changed. 

In more detailed models, the common procedure to investigate a possible multiple equilibrium 
regime is to slowly increase the freshwater forcing in the Northern Atlantic until a critical threshold is crossed: the stable 
AMOC state ceases to exist and the circulation collapses to the other stable state without an active AMOC in a bifurcation 
tipping event. Subsequently the forcing is reduced again until the collapsed state no longer exists and the AMOC 
recovers. The interval in forcing conditions between  collapse and recovery is the AMOC hysteresis width and a 
nonzero width indicates the existence of multiple equilibria.  
Such AMOC hysteresis  has been 
found in Earth system Models of  Intermediate Complexity \protect\citep{rahmstorf2005thermohaline, lenton2007effects} and 
global ocean  models \protect\citep{hofmann2009stability}.  Full AMOC hysteresis simulations have also been performed   with a 
low resolution atmosphere-ocean coupled general circulation  model (FAMOUS)  \protect\citep{hawkins2011bistability},  and 
 in a state-of-the-art global climate model (CESM) \protect\citep{van2023asymmetry}.

The AMOC is identified as a tipping element in the climate system \protect\citep{lenton2008tipping}. Based on early warning signals determined from historical AMOC reconstruction data \protect\citep{Caesar2018}, the AMOC is believed to be heading towards the bifurcation point \protect\citep{Boers2021}. Although estimates have recently been provided when such a bifurcation point would be reached \protect\citep{Ditlevsen2023}, many uncertainties remain and the present-day AMOC state may be far from it. Therefore, what is more dangerous is a noise-induced transition, where the transition does not happen because a certain critical threshold was crossed, but due to the fast variability in the forcing ('noise'). Not only are there no reliable early warning signals for a noise-induced transition as opposed to a bifurcation-induced one \protect\citep{ditlevsen2010tipping}, but more so that the latter can only occur close to the bifurcation point whereas the former can take place as long as the AMOC is multi-stable. This is arguably the case for the present-day AMOC. Based on observations of the AMOC induced freshwater transport at the southern boundary of the Atlantic, the present-day AMOC is thought to be in a multiple equilibrium regime \protect\citep{Weijer2019}. The reliability of this transport as an indicator for multi-stability is however contentious \protect\citep{jackson2018hysteresis}. Several stochastic models of the AMOC have been developed with stochastic freshwater forcing and the statistics of their noise-induced transitions have been analysed  \protect\citep{cessi1994simple, timmermann2000noise} and recently even a noise-induced long-time weakening of the AMOC has been found in a global model \protect\citep{romanou2023stochastic}.

However, as there is no observational evidence for an AMOC transition over the historical period, a noise-induced transition is expected to be a rare event and simple Monte Carlo techniques are unsuitable for such low-probability events. Rare event techniques have been used to determine such transitions probabilities in box models \protect\citep{castellana2019transition, jacques2024estimation}. However, apart from such probabilities, information on the most likely path of such an AMOC transition is also desired. From this, likely new observable early warning signals can be determined of an AMOC collapse. 

In this paper, we use a technique from Large Deviation Theory (LDT) to directly compute the most likely transition path in the low-noise limit of a stochastic model,  i.e. the instanton. This is achieved by minimizing the Freidlin-Wentzell action \protect\citep{freidlin1998random}, transforming a rare transition sampling problem into a deterministic one. Minimizing this action entails that we maximize the probability of the applied noise forcing under the constraints that a transition occurs. The resulting path (the instanton) allows us to analyse the mechanisms of a noise-induced transition. Note that the forcing implies the path (and vice-versa) so with the most likely path comes the optimal forcing. This associated forcing is  just sufficient enough to attain the rare event. In other scientific fields  instantons have already been 
used  to analyze transitions \protect\citep{grafke2015instanton, grafke2019numerical, woillez2020instantons, schorlepp2022spontaneous, blyuss2023sex}. We apply the technique here to compute the instanton 
for an AMOC collapse and recovery in the box model  from \protect\citet{wood2019observable},  
where stochastic freshwater forcing is added. We will present the paths for an AMOC collapse and recovery that have the most probable freshwater forcing needed to achieve these transitions.

In section \protect\ref{sec:ldt}, a brief introduction to the Freidlin-Wentzell theory of Large Deviations is provided and the 
algorithm used to compute the instanton is presented. Then, section \protect\ref{sec:wood} discusses the stochastic 
version of the box model from \protect\citet{wood2019observable} together with the assumptions on the applied freshwater noise. Section 
\protect\ref{sec:trs} presents results on the instantons and the accompanying optimal forcings found for an AMOC collapse and recovery for the stochastic box model.  In section \protect\ref{sec:hoos}, the distribution of this stochastic freshwater forcing over the boxes in the model is altered to study the effects of  different freshwater forcing on the most likely AMOC collapse and recovery. Furthermore, the instanton trajectories are compared to the trajectories resulting from a standard deterministic hosing experiment. Finally,  in section \protect\ref{sec:con}, a summary and discussion of the results follow.

\section{Theory} \label{sec:ldt}
\subsection{Freidlin-Wentzell Theory of Large Deviations}
Consider a dynamical system with state vector  $X_t$ in $\R^d$, forced by small random perturbations which are white in time and additive Gaussian. The noise amplitude scales with smallness-parameter $\epsilon$. The system can then be described by the stochastic differential equation (SDE):
\begin{equation}\label{eq:genSDE}
    dX_t = f(X_t)\,dt + \sqrt{\epsilon}\sigma\,dW_t,\quad\quad t\geq0 , 
\end{equation}
where $f:\,\R^d\to\R^d$ is the deterministic drift, $a = \sigma\sigma^T$ the noise covariance with $\sigma\in\R^{d\times k}$, and $W_t$ a $k$-dimensional Wiener process. For simplicity we take $a$ to be independent of the system's state  (although this theory can be generalized to include multiplicative noise), but it is not assumed that $a$ is invertible, i.e. the noise can be degenerate. Of interest are situations where the stochastic process \protect\eqref{eq:genSDE} realizes a certain event, say a transition within time $T$ starting at $X_0 = x$ and ending at $X_T = y$ for certain points in phase space $x,y\in \R^d$ and time $T$. These events might be impossible in the deterministic setting ($\epsilon = 0$) but can occur in the presence of noise ($\epsilon >0$), although they become increasingly rare in the low-noise limit ($\epsilon\to0$).

LDT provides the rate at which this probability decays to zero with $\epsilon$. The probability of observing a sample path close to a function $\phi:\,[0,T]\to\R^d\,\,\big(t\mapsto\phi(t)\big)$ behaves as:
\begin{equation}\label{eq:LDP}
    \Prb\bigg[\sup_{t\in[0,T]}\|X_t-\phi(t)\|<\delta\bigg]\asymp\exp\big[-S_T[\phi]/\epsilon\big] 
\end{equation}
for sufficiently small $\delta>0$, and where $\asymp$ denotes the log-asymptotic equivalence. 
The functional $S_T[\phi]$ is the Freidlin-Wentzell action and is defined as
\begin{equation}\label{eq:action}
    S_T[\phi] = \frac{1}{2}\int_0^T\langle\dot{\phi}(t)-f(\phi(t)),\,a^{-1}(\dot{\phi}(t)-f(\phi(t)))\rangle\,dt 
\end{equation}
if the integral converges, otherwise $S_T[\phi] = \infty$. Here, $\langle\cdot,\cdot\rangle$ is the standard Euclidean inner product which induces the norm $\|\cdot\|$, and $a^{-1}$ the inverse of the covariance matrix \protect\citep{freidlin1998random}. More technical constraints are that $f(\phi)$ needs to be Lipschitz-continuous in $\phi$, and the path $\phi(t)$ needs to be absolutely continuous in time $t$. Moreover, if $a$ is singular then $a^{-1}$ can be replaced by the Moore-Penrose inverse provided a positive-definite $a^{1/2}$ exists, and provided $\dot{\phi}-f(\phi)$ is in the image of $a$ \protect\cite[]{puhalskii2004some}. 

The important consequence of \protect\eqref{eq:LDP} is that in the limit of $\epsilon\to0$ the trajectory $\phi^*(t)$ with the smallest action becomes the least unlikely trajectory to realize the rare event, and all sample paths conditioned on the rare event ($\phi(0) = x,\,\phi(T) = y$) will concentrate around $\phi^*$. More precisely:
\begin{equation}\label{eq:min}
    \phi^*(t) = \argmin_{\{\phi(0) = x,\,\phi(T)=y\}}S_T[\phi] 
\end{equation}
is the instanton (or maximum likelihood pathway), and for $\delta >0$ sufficiently small
\begin{equation*}
    \lim_{\epsilon\to0}\Prb\bigg[\sup_{t\in[0,T]}\|X_t-\phi^*(t)\|<\delta\,\Big|X_0 = x,\,X_T = y\bigg] = 1.
\end{equation*}
So the objective here is to compute the instanton $\phi^*$ i.e. the minimizing path of the 
Freidlin-Wentzell action functional under the constraints that this is a transition path. Simply speaking, it is the path whose associated forcing is the most likely to occur of all possible forcings that cause a transition in a time $T$.

\subsection{Hamiltonian Principle}
The minimization problem \protect\eqref{eq:min} is a common problem within classical mechanics \protect\citep{taylor2005classical} and so similar methods can be applied. The Lagrangian of \protect\eqref{eq:min} is
\begin{equation*}
    \mathcal{L}(\phi, \dot{\phi}) = \frac{1}{2}\langle\dot{\phi}(t)-f(\phi(t)),\,a^{-1}(\dot{\phi}(t)-f(\phi(t)))\rangle
\end{equation*}
and assuming convexity of $\mathcal{L}(\phi, \dot{\phi})$ in $\dot{\phi}$, a Hamiltonian can be formulated by taking its Fenchel-Legendre transform \protect\citep{grafke2019numerical}
\begin{equation*}
    \mathcal{H}(\phi,\theta) = \sup_y\big(
    \langle\theta,y\rangle-\mathcal{L}(\phi,y)\big) , 
\end{equation*}
where $\theta$ is the conjugate momentum defined as
\begin{equation*}
    \theta = \derp{\mathcal{L}(\phi, \dot{\phi})}{\dot{\phi}}.
\end{equation*}
Hence the Hamiltonian is
\begin{equation}
    \mathcal{H}(\phi,\theta) = \langle f(\phi),\theta\rangle + \frac{1}{2}\langle\theta,a\theta\rangle
        \label{e:IHamilton}     
\end{equation}
and so the Hamiltonian equations of motion, i.e. the instanton equations follow:
\begin{align}
    \begin{cases}
        \dot{\phi} &= \nabla_\theta\mathcal{H(\phi,\theta)} = f(\phi) + a\theta\\
        \dot{\theta} &= -\nabla_\phi\mathcal{H(\phi,\theta)} = -(\nabla f(\phi))^T\theta
    \end{cases}
    \label{e:Instantoneq} 
\end{align}
with boundary conditions $\phi(0) = x$ and $\phi(T) = y$. Solving these equations yields the instanton $\phi^*(t)$. Note that for this path the Hamiltonian is conserved. An advantage of this approach is that the covariance matrix $a$ does not need to be inverted. A disadvantage is that for one first-order equation two boundary conditions are known, while for the conjugated equation there are none. A naive solution would be to guess $\theta(0)$ such that $\phi(T)$ is close to $y$ i.e. a shooting method, which is unpractical in higher dimensions. So to solve \protect\eqref{e:Instantoneq} we use the Augmented Lagrangian 
method \protect\citep{hestenes1969multiplier, schorlepp2022spontaneous}, see Appendix A for details. 

\section{Model} \label{sec:wood}
\begin{figure}[h!]
    \centering
    \includegraphics[scale = 0.3]{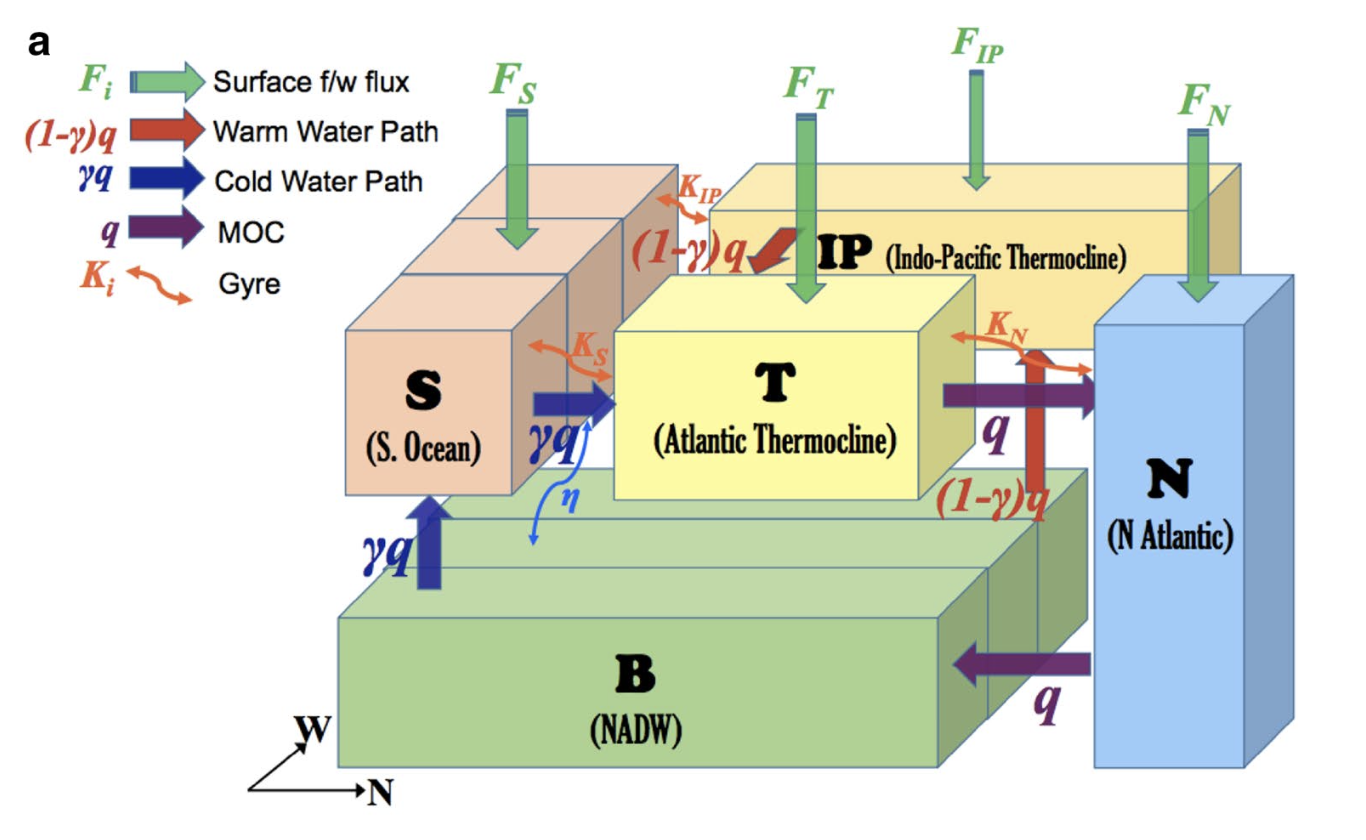}
    \caption{Sketch of the box model, from \protect\cite{wood2019observable}.}
    \label{fig:cimamodel}
\end{figure}

The ocean model used was originally introduced by \protect\cite{wood2019observable}, and is chosen since it is calibrated to the FAMOUS-model \protect\citep{smith2008description, smith2012famous}). They show that this simple box model encapsulates the physics of the more complicated AOGCM. It represents the World Ocean in five boxes, see figure \protect\ref{fig:cimamodel}. Each box represents large contiguous regions of the global ocean, corresponding to large-scale water mass structures. The \textit{T} box represents the Atlantic thermocline; the \textit{N} box is the North Atlantic Deep Water (NADW) formation region and Artic region; the \textit{B} box is the southward propagating NADW and its upwelling in the Southern Ocean as Circumpolar Deep Water; the \textit{S} box is the fresh Southern Ocean near-surface waters and their return into the Atlantic as Antarctic Intermediate Water, and lastly the \textit{IP} box represents the Indo-Pacific thermocline. The calibration entails that the box model reproduces the decadal means of salinities, temperatures, volumes and transports of these mean water masses observed in the stable pre-industrial climate state of the AOGCM. The boxes have salinities denoted by $S_N$, $S_T$, $S_S$, $S_B$ and $S_{IP}$ respectively, along with fixed volumes $V_N$, $V_T$, $V_S$, $V_B$ and $V_{IP}$. 

\subsection{Formulation}
It is assumed that density variations in the boxes are solely due to changes in the boxes' salinity and in the temperature of the northern box. The other boxes' temperatures are taken to be constant. The following volume fluxes occur between the boxes: downwelling in the North Atlantic (and hence AMOC strength) $q$ that transports salt between the \textit{T}, \textit{N} and \textit{B} box, a cold water path (CWP) that returns a fraction $\gamma$ of the AMOC from the \textit{B} box via the \textit{S} box to the \textit{T} box, representing the return flow via the Drake Passage, and a warm water path (WWP) that returns the AMOC's remaining fraction $1-\gamma$ from the \textit{B} box via the \textit{IP} box to the \textit{T} box, representing the return flow via the Agulhas leakage. It is assumed that the overturning circulation is linear dependent on the density difference of the \textit{N} and \textit{S} boxes:
\begin{align*}
    q = \lambda(\alpha(T_S-T_N)+\beta(S_N-S_S)) , 
\end{align*}
with $\alpha$ and $\beta$ the thermal and haline compressibility, $\lambda$ a hydraulic coefficient, and 
$T_S$ the fixed mean temperature of the \textit{S} box. Additionally,  $T_N$ is the mean temperature of the northern box obeying 
\begin{align*}
    T_N = \mu q + T_0.
\end{align*}
Here $T_0$ is a fixed baseline temperature, and $\mu>0$ the fixed heat transport coefficient. This represents the northward transport of heat by the overturning circulation. Combining the latter two equations yields:
\begin{align*}
    q &= \kappa (\alpha(T_S-T_0)+\beta(S_N-S_S)) , \\
    &\text{ where }\kappa = \frac{\lambda}{1+\alpha \lambda \mu }
\end{align*}
eliminating $T_N$ from the equations and hence the state of the dynamical model is completely characterized by the boxes' five salinities.

The other salinity transports are due to the wind-driven subtropical gyres, captured by the coefficients $K_N$, $K_S$ and $K_{IP}$ for the exchange between \textit{T} and \textit{N}, \textit{T} and \textit{S}, and \textit{IP} and \textit{S} respectively. Furthermore, the coefficient $\eta$ captures the mixing between \textit{B} and \textit{S}. The atmospheric fluxes are denoted by $F_i$ for the flux into box $i$. 

The stochastic extension of the  \protect\cite{wood2019observable} model is given by the equations: 
\begin{subequations}
\begin{align}
    V_N\der{S_N}{t} &= q\Big(\theta(q)(S_T-S_N)-\theta(-q)(S_B-S_N)\Big)\nonumber\\
    &+ K_N(S_T-S_N) - F_NS_0 + A_NS_0\Tilde{\eta}\label{eq:Wooda}\\
    V_T\der{S_T}{t} &= q\Big(\theta(q)(\gamma S_S+(1-\gamma)S_{IP}-S_T)\nonumber\\
    &-\theta(-q)(S_N-S_T)\Big) + K_S(S_S-S_T)\nonumber\\
    &- F_TS_0 + A_TS_0\Tilde{\eta}\label{eq:Woodb}\\
    V_S\der{S_S}{t} &= \gamma q\Big(\theta(q)(S_B-S_S)-\theta(-q)(S_T-S_S)\Big)\nonumber\\ 
    &+ K_{IP}(S_{IP}-S_S) + K_S(S_T-S_S)\nonumber\\
    &+\eta(S_B-S_S)- F_SS_0 + A_SS_0\Tilde{\eta}\label{eq:Woodc}\\
    V_{IP}\der{S_{IP}}{t} &= (1-\gamma) q\Big(\theta(q)(S_B-S_{IP})\nonumber\\ 
    &-\theta(-q)(S_T-S_{IP})\Big)+ K_{IP}(S_S-S_{IP})\nonumber\\
    &- F_{IP}S_0 + A_{IP}S_0\Tilde{\eta}\label{eq:Woodd}\\
    V_B\der{S_B}{t} &= q\Big(\theta(q)(S_N-S_B)\nonumber\\
    &-\theta(-q)(\gamma S_S + (1-\gamma)S_{IP} - S_B)\Big)\nonumber\\
    &+\eta(S_S-S_B)\label{eq:Woode}
\end{align}    
\end{subequations}
where $S_0$ is a baseline salinity value and $\theta(x)$ the Heaviside function.
The stochastic freshwater flux $\Tilde{\eta}(t)$ is modelled as 
a zero-mean Gaussian white noise process with correlation function $\E(\Tilde{\eta}(t)\Tilde{\eta}(t')) = 
\nu\delta(t-t')$. The parameters $A_i$ indicate what fraction of  $\Tilde{\eta}$ 
enters or exits box $i$. 

The total salinity $C$ is 
\begin{align}
    C = V_NS_N + V_TS_T + V_SS_S + V_{IP}S_{IP} + V_BS_B,
\label{e:Scons}
\end{align}
and using (7), it follows that 
\begin{align*}
    \der{C}{t} &= -(F_N+F_T+F_S+F_{IP})S_0 , \\
    &+ (A_N+A_T+A_S+A_{IP})S_0\Tilde{\eta} . 
\end{align*}
The total salinity is conserved, so it is demanded that
\begin{align*}
    F_N+F_T+F_S+F_{IP} &= 0 ,  \\
    A_N+A_T+A_S+A_{IP} &= 0 . 
\end{align*}
Using (\protect\ref{e:Scons}), the differential equation \protect\eqref{eq:Woode} is replaced by:
\begin{align*}
    S_B &= \frac{1}{V_B}\big(C -V_NS_N -V_TS_T -V_SS_S -V_{IP}S_{IP}\big)
\end{align*}
with $C$ the basin's total (fixed) salinity. 

\subsection{Non-dimensional equations}
The non-dimensional salinities $\phi_i$ of the boxes are introduced, together with non-dimensionalized volumes $V_i$:
\begin{align*}
    &S_N = S_0\phi_1 & V_N = V_0 V_1\\
    &S_T = S_0\phi_2 & V_T = V_0 V_2\\
    &S_S = S_0\phi_3 & V_S = V_0 V_3\\
    &S_{IP} = S_0\phi_4 & V_{IP} = V_0 V_4\\
    &S_B = S_0\phi_5 & V_B = V_0 V_5
\end{align*}
and furthermore time $t$ is non-dimensionalized as $t = t_d\tau$, with notation $\dot{\phi}_i = \der{\phi_i}{\tau}$. 
The non-dimensional model equations then become:
\begin{subequations}
\begin{align}
    \dot{\phi}_1 &= \frac{t_d}{V_0}\frac{1}{V_1}\Big(q[\theta(q)(\phi_2-\phi_1)-\theta(-q)(\phi_5-\phi_1)]\nonumber\\
    &+ K_N(\phi_2-\phi_1)-F_N\Big)+ \frac{t_d}{V_0}\frac{A_{N}}{V_1}\sqrt{\nu}\zeta\label{eq:nondima}\\
    \dot{\phi}_2 &= \frac{t_d}{V_0}\frac{1}{V_2}\Big(q[\theta(q)(\gamma\phi_3+(1-\gamma)\phi_4-\phi_2)\nonumber\\
    &-\theta(-q)(\phi_1-\phi_2)]+ K_S(\phi_3-\phi_2)\nonumber\\
    &+K_N(\phi_1-\phi_2)-F_T\Big)+ \frac{t_d}{V_0}\frac{A_{T}}{V_2}\sqrt{\nu}\zeta\label{eq:nondimb}\\
    \dot{\phi}_3 &=\frac{t_d}{V_0}\frac{1}{V_3}\Big(\gamma q[\theta(q)(\phi_5-\phi_3)-\theta(-q)(\phi_2-\phi_3)]\nonumber\\
    &+ K_{IP}(\phi_4-\phi_3) +K_S(\phi_2-\phi_3)+\eta(\phi_5-\phi_3)-F_S\Big)\nonumber\\
    &+ \frac{t_d}{V_0}\frac{A_{S}}{V_3}\sqrt{\nu}\zeta\label{eq:nondimc}\\
    \dot{\phi}_4 &=\frac{t_d}{V_0}\frac{1}{V_4}\Big((1-\gamma)q[\theta(q)(\phi_5-\phi_4)-\theta(-q)(\phi_2-\phi_4)]\nonumber\\
    &+ K_{IP}(\phi_3-\phi_4) -F_{IP}\Big) + \frac{t_d}{V_0}\frac{A_{IP}}{V_4}\sqrt{\nu}\zeta\label{eq:nondimd}
\end{align}
\end{subequations}
where $\zeta(t)$ is a unit-variance white noise process and the (dimensional) AMOC 
strength is written as 
\begin{align}\label{eq:q}
    q &= \kappa(\alpha(T_S-T_0) + \beta S_0(\phi_1-\phi_3))
\end{align}
and non-dimensionalized $S_B$ follows from salinity conservation
\begin{align*}
    \phi_5 &= \Big(\frac{C}{V_0S_0}-V_1\phi_1-V_2\phi_2-V_3\phi_3-V_4\phi_4\Big)/V_5.
\end{align*}
From now on $t$ is again used, but now  to denote the non-dimensional time. 

The system of non-dimensionalized equations is rewritten as an It\^o
stochastic differential equation \protect\citep{mikosch1998elementary}:
\begin{equation}\label{eq:model}
    d\phi_t = f(\phi_t)\,dt + \sqrt{\epsilon}\sigma dW_t
\end{equation}
where $\phi_t = \big(\phi_1(t),\phi_2(t),\phi_3(t),\phi_4(t)\big)^T$ is the state vector, and the 
deterministic parts of the evolution equations are $f(\phi) = \big(f_1(\phi),f_2(\phi),f_3(\phi),f_4(\phi)\big)^T$. 
Furthermore,  the noise amplitude is given by $\sqrt{\epsilon} = \frac{t_d}{V_0}\sqrt{\nu}$ and 
\begin{equation*}
    \sigma = (A_N/V_1,A_T/V_2,A_S/V_3,A_{IP}/V_4)^T
\end{equation*}
acts on the 1-dimensional Wiener process.
The system has degenerate noise: one degree of stochastic freedom is acting on 
the process $\phi_t\in\R^4$.

The parameter values ($V_i$'s, $F_i$'s, $K_i$'s, $T_S$, $T_0$, $\mu$, $\lambda$, $\eta$ and $\gamma$) are 
taken from the calibration to the FAMOUS\_B-model\protect\citep{wood2019observable}, see 
Table \protect\ref{tab:par} in Appendix B. The total salinity $C$ is found using the salinity 
values from \protect\citet{alkhayuon2019basin}.  For these values the deterministic model has two stationary stable points, an ON and OFF-state of the AMOC ($q \lessgtr 0$) and an unstable saddle state \protect\citep{alkhayuon2019basin}. In the ON-state ($q>0$) downwelling occurs in the north, and the resulting flow  flows southward along the bottom 
before it returns to the Atlantic thermocline via either the Southern Ocean or the Indo-Pacific  thermocline. In the 
OFF-state these flows are reversed, and upwelling occurs in the Northern Atlantic ($q<0$). So states $\phi_{\text{ON}}$ and $\phi_{\text{OFF}}$ represent an active and collapsed AMOC respectively. They are 
found numerically using a root finding function, the dogleg method \protect\citep{nocedal1999numerical}, 
to $f:\,\R^4\to\R^4$.

\subsection{Instanton Equations}
Recall that we want to compute the instantons for an AMOC collapse ($\phi_{\text{ON}}\to\phi_{\text{OFF}}$) and recovery ($\phi_{\text{OFF}}\to\phi_{\text{ON}}$) in model \protect\eqref{eq:model}. 
So we need $f(\phi)$ to be Lipschitz-continuous in $\phi$. We did this,  by approximating the Heaviside function by $\theta(x) \approx (\erf{(x/\epsilon_\theta)}+1)/2$, where $\erf$ is the error function,  for small $\epsilon_\theta = 10^{-10}$. The error caused by this approximation is assumed to be negligible. 

Two assumptions were made of the freshwater noise:(i) the noise  is white in time, \'and (ii) this 
noise is  small. To justify (ii), estimating the noise amplitude can be done in a similar way as \protect\citet{castellana2019transition}. The parameter $\epsilon$ is estimated by using the annual $P-E$ (precipitation minus evaporation) time series in the Atlantic Ocean, assuming that this represents the inter annual variability of freshwater perturbations relevant for the AMOC strength. Using the same time series, the ERA-Interim Archive at ECMWF \protect\citep{dee2011era}, it is found for its variance that $\sqrt{\nu}\sim 0.0175\pm0.0025$Sv. Hence $\epsilon \sim 0.0055\pm0.0008$. This is in the regime of small noise, as for $\sqrt{\nu}\sim0.07$ Sv typically no transition occurs even after $10^6$ years, while a transition usually does not take longer than $35$ years. Assumption (i) is harder to justify as the annual freshwater perturbation series might be auto-correlated, especially under climate change. However, the white noise assumption is still made to keep the model analysis relatively simple. 

Applying the equations from section \protect\ref{sec:ldt} to the stochastic box model yields the Lagrangian
\begin{equation*}
    \mathcal{L}(\phi,\dot{\phi}) = \frac{1}{2\|\sigma\|^4}\sum_{i,j =1}^4\sigma_i\sigma_j\big(\dot{\phi}_i-f_i(\phi)\big)\big(\dot{\phi}_j-f_j(\phi)\big)
\end{equation*}
with conjugate momentum $\theta:\,[0,T]\to\R^4$
\begin{equation*}
    \theta_i(t) = \frac{\sigma_i}{\|\sigma\|^4}\sum_{j=1}^4\sigma_j\big(\dot{\phi}_j(t)-f_j(\phi_j(t))\big)\quad\text{ for }1\leq i\leq4
\end{equation*}
and hence the Hamiltonian for this minimization problem is 
\begin{equation*}
    \mathcal{H}(\phi,\theta) = \langle f(\phi), \theta\rangle + \frac{1}{2}\|\sigma\theta\|^2.
\end{equation*}
Therefore the equations for the instanton from $\phi_{\text{start}}$ to $\phi_{\text{end}}$ within a time interval $[0,T]$ are:
\begin{align}\label{eq:inst}
    \begin{cases}
        \dot{\phi}(t) &= f(\phi(t)) + \sigma\langle\sigma, \theta(t)\rangle\\
        \dot{\theta}(t) &= -\big(\nabla f(\phi(t))\big)^T\theta(t)\\
        &\text{with }\phi(0) = \phi_{\text{start}}\,\text{ and }\phi(T) = \phi_{\text{end}}.
    \end{cases}
\end{align}

As the stochastic forcing is just a one-dimensional Wiener process, all deviations from the deterministic system are the same in every direction up to a scaling factor $\sigma_i^{-1}$, and one can define:
\begin{align*}
    \xi(t) &= \big(\dot{\phi}_i(t) - f_i(\phi(t))\big)/\sigma_i\quad\text{ for }1\leq i\leq 4\\
    &= \langle\sigma, \theta(t)\rangle
\end{align*}
which reduces the previously mentioned relationships to:
\begin{subequations}
 \begin{align}
    \mathcal{L}(\xi) &= \frac{1}{2}\xi^2\\
    \mathcal{H}(\phi,\xi) &= \frac{\langle f(\phi),\sigma\rangle}{\|\sigma\|^2}\xi + \frac{1}{2}\xi^2\label{e:Ham_Mod}.
 \end{align}   
\end{subequations}
The Lagrangian shows trivially that the most likely path is the one that minimizes the absolute total applied freshwater noise, while the Hamiltonian implies that to any freshwater perturbation the system $\phi$ responds with salinity changes in all boxes with proportionality factors $\sigma_i/\|\sigma\|^2$. We will obtain the path $\phi$ and accompanying forcing $\xi$ that minimize this Lagrangian and do not vary the Hamiltonian -- provided that this path is a transition.

\section{Results: AMOC Transitions} \label{sec:trs}
\subsection{Instantons}
The instanton for an AMOC collapse from $\phi_{\text{ON}}$ to $\phi_{\text{OFF}}$ and its reverse (the `anti-instanton') are computed for the  noise parameters $\big(A_N,A_T,A_S,A_{IP}\big)^T = \big(0.070, 0.752, -0.257, -0.565\big)$ which follow from the same calibration to the FAMOUS\_B model \protect\citep{wood2019observable}. So freshwater goes into boxes \textit{N} and \textit{T}, and as compensation freshwater is extracted from boxes \textit{S} and \textit{IP}. This conserves the total salinity and so $\phi_{\text{ON}}$ and $\phi_{\text{OFF}}$ do not change during the hosing process. The equations \protect\eqref{eq:inst} are solved using the discussed Augmented Lagrangian method, where explicit Euler is used for time integration with time step $\Delta t = 0.05$. The instantons $\phi(t)$ are terminated once $\|\phi(T)-\phi_{\text{end}}\| < 10^{-5}$. For different values of $30<T<100$ the trajectory of the instanton is just translated in time without any significant change in behavior. Moreover the average times of a realization to transit from the ON-basin (defined by $q>14$ Sv) to the OFF-basin (defined by $q<-4.5$ Sv) and vice versa level off for smaller noise levels (tested down to $\epsilon \sim 0.035$) at $T = 17$ and $T = 32$ for the collapse and recovery respectively. Hence  $T = 32$ is sufficient to capture the relevant instanton behavior for both transition directions.  

In the limit of large $T$ the instantons are expected to cross from one basin to the other via the saddle present in the model. We do not consider these instantons to be representable of a typical transition, since the realizations here tend to avoid the saddle, which is an issue discussed in \cite{borner2023saddle}.

\begin{figure}
    \centering
    \includegraphics[width = 19pc]{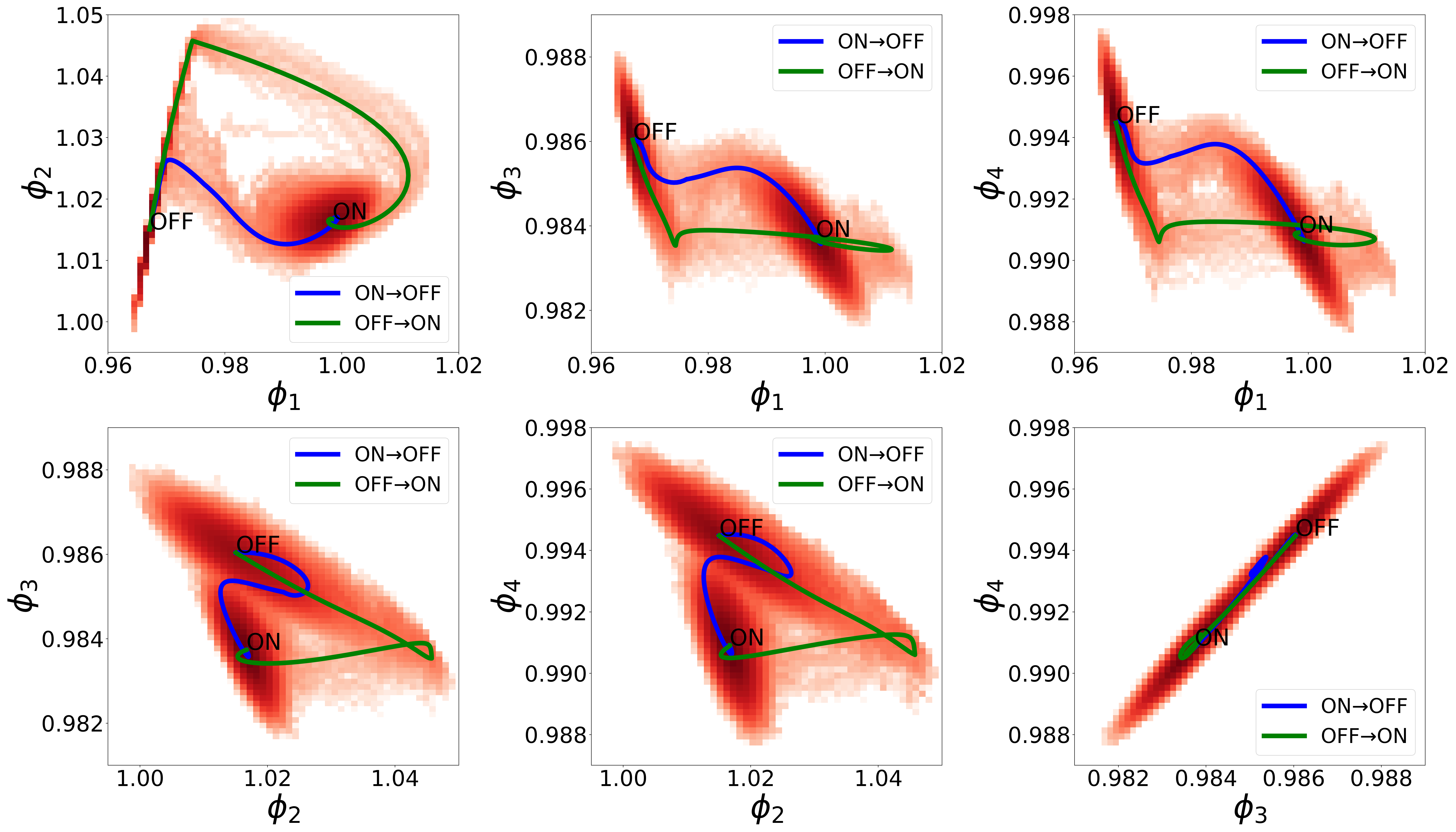}
    \caption{Instantons from $\phi_{\text{ON}}$ to $\phi_{\text{OFF}}$ (blue) and vice versa (anti-instantons, green) in several salinity spaces $(\phi_i,\phi_j)$, together with a histogram of multiple realizations of the model for noise level $\epsilon \sim 0.035$ \big($\sqrt{\nu}\sim0.11$Sv\big) (red), using a logarithmic scale.}
    \label{fig:overviewONOFF}
\end{figure}

In addition to  the instanton trajectory $\phi(t)$, the associated optimal freshwater noise forcing $\xi(t)$ also follows from the optimization procedure \eqref{eq:inst} and is analyzed below. In Fig.~\protect\ref{fig:overviewONOFF}, 
the resulting instantons are shown in several projected phase spaces $(\phi_i,\phi_j)$ together with multiple realized transitions 
of the model. It  shows that the instantons lie at the centre of these transitions, as they should be according to theory, and 
this serves as their verification.  We will analyse the mechanisms underlying  the behavior of these  instantons in the following 
subsections. 

\subsection{Dynamics of Collapse \& Recovery}
\begin{figure*}
    \centering
    \includegraphics[width = 39pc]{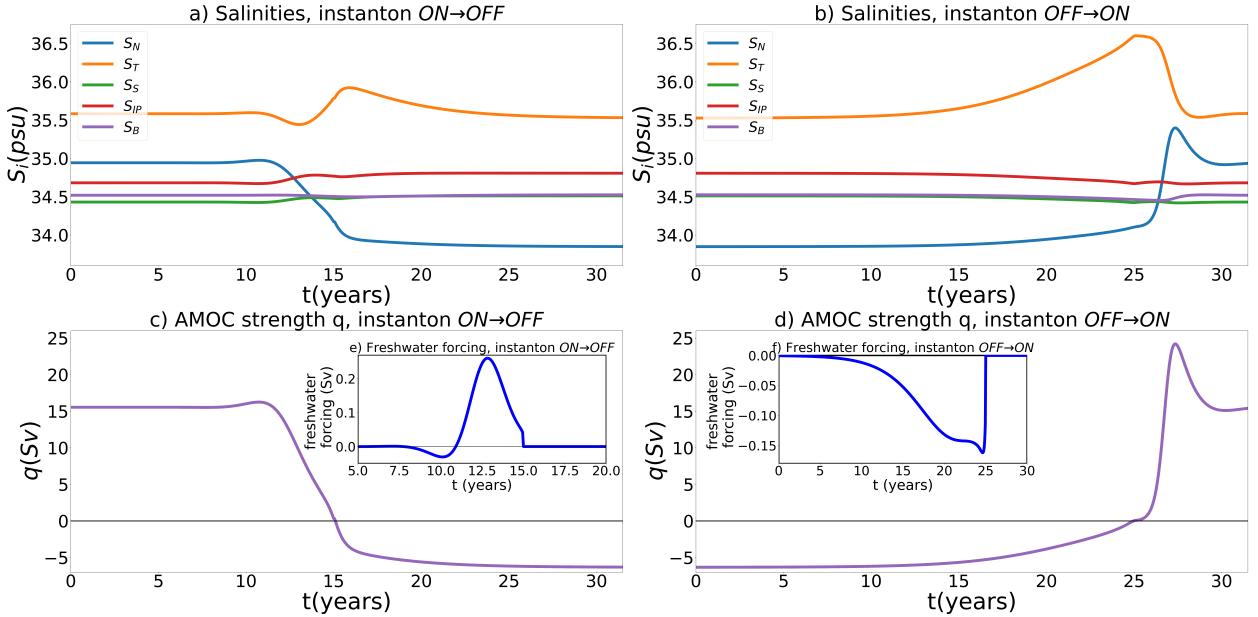}
    \caption{Instantons from the ON-state to the OFF-state (left) and reverse (right) shown as the five salinities (top) and AMOC strength (bottom), with the optimal freshwater forcing as insets. The salinities $S_N$, $S_T$, $S_S$, $S_{IP}$ and $S_B$ are indicated by blue, yellow, green, red and purple respectively.}
    \label{fig:overview_every}
\end{figure*}

In Fig.~\protect\ref{fig:overview_every} the dimensional instantons of the collapse and recovery are shown together with their AMOC strength $q$ for a time frame $t\in[0,32]$ years. The insets show the accompanying dimensional stochastic freshwater forcings  into boxes \textit{N} and \textit{T}, which are computed as $-\frac{V_0}{t_d}\big(V_1\sigma_1+V_2\sigma_2\big)\xi(t)$. A positive sign indicates net freshwater going into these boxes. Mainly the salinities in the \textit{N} and \textit{T} boxes change, while those in the other boxes only vary marginally. \protect\citet{wood2019observable} observed this already, and suggested a possible reduction of the model without \textit{S} and \textit{IP}. In those boxes advection by the density-induced flow plays a lesser role, and their salinities are mainly controlled by the atmospheric fluxes and the Indo-Pacific gyre transport. Hence the tipping of the AMOC affects these boxes to a lesser extent than the 
\textit{N} and \textit{T} boxes. 

The dynamics of the AMOC collapse and recovery can be captured by a reduced model. We assume 
$\phi_3$ (i.e. $S_S$)  is constant and that $A_N$ is negligible. The former assumption is justified as it can be deduced from figure \ref{fig:overview_every}, while the latter is justified since the hosing transport into box \textit{N} is several orders of magnitude smaller than the other transports. Now for $q>0$ it holds that:
\begin{align*}
    \dot{q} &= \kappa\beta S_0\dot{\phi}_1\\
    &= \kappa\beta S_0\frac{t_d}{V_0V_1}\big[(K_N+q)(\phi_2-\phi_1)-F_N\big]
\end{align*}
then substituting the relation $\phi_1(q)$ derived directly from \protect\eqref{eq:q} yields
\begin{align}\label{eq:qdoton}
    \dot{q}&_{\text{ON}}(q,\phi_2) = A_{\text{ON}}q^2 + B_{\text{ON}}q + C_{\text{ON}}\\
    &\text{with}\nonumber\\
    &A_{\text{ON}} = -\frac{t_d}{V_0V_1}\nonumber\\
    &B_{\text{ON}} = \frac{t_d}{V_0V_1}\big(\kappa\beta S_0(\phi_2-\phi_3) - K_N +\kappa\alpha(T_S-T_0)\big)\nonumber\\
    &C_{\text{ON}} = \frac{t_d}{V_0V_1}\kappa\beta S_0\big(K_N(\phi_2+\frac{\alpha(T_S-T_0)}{\beta S_0}-\phi_3)-F_N\big).\nonumber
\end{align}
A similar derivation can be done for the AMOC dynamics as $q<0$ which leads to
\begin{align}\label{eq:qdotoff}
    \dot{q}&_{\text{OFF}}(q,\phi_2) = A_{\text{OFF}}q^2 + B_{\text{OFF}}q + C_{\text{OFF}}\\
    &\text{with}\nonumber\\
    &A_{\text{OFF}} = \frac{t_d}{V_0V_1}\nonumber\\
    &B_{\text{OFF}} = \frac{t_d}{V_0V_1}\big(\kappa\beta S_0(\phi_3-\phi_5) - K_N -\kappa\alpha(T_S-T_0)\big)\nonumber\\
    &C_{\text{OFF}} = \frac{t_d}{V_0V_1}\kappa\beta S_0\big(K_N(\phi_2+\frac{\alpha(T_S-T_0)}{\beta S_0}-\phi_3)-F_N\big).\nonumber
\end{align}
The evolution of  $\dot{q}(q,\phi_2)$ along  the instantons is depicted in 
Fig.~\protect\ref{fig:parabola}.

\subsubsection{Collapse}
Initially the system is in equilibrium $\phi_{\text{ON}}$, which is a stable root of $\dot{q}_{\text{ON}}(q,\phi_2) = 0$ (Fig.~\protect\ref{fig:parabola}a). This right root is stable, since a small increase (decrease) in $q$ would cause the \textit{N} box to become saltier (fresher), which is mitigated by a decrease (an increase) in Northern gyre transport that in turns freshens (salinifies) the Northern box, which finally weakens (strengthens) the AMOC. This stable root will continue to exist as long as the discriminant of $\dot{q}_{\text{ON}}(q,\phi_2)$ stays positive, i.e. $D_{\text{ON}} = B_{\text{ON}}^2-4A_{\text{ON}}C_{\text{ON}} >0$. It can be derived that this condition is equivalent to 
\begin{align}\label{eq:stabroot}
    \phi_2-\phi_3 &> r(T_S-T_0)-k_N + 2\sqrt{f_N} \approx 0.0494\\
    \text{or }\phi_2-\phi_3 &< r(T_S-T_0)-k_N - 2\sqrt{f_N} \approx -0.0477\nonumber\\
    &\text{ where }k_N = \frac{K_N}{\kappa\beta S_0},\, r = \frac{\alpha}{\beta S_0}\text{ and }f_N = \frac{F_N}{\kappa \beta S_0}\nonumber
\end{align}
where only the first inequality is considered as $\phi_2-\phi_3<0$ does not happen. Looking at this inequality it follows that a larger Northern gyre strength, a warmer Southern box or a smaller atmospheric flux into box \textit{N} would all enlarge this range for $\phi_2-\phi_3$. This follows directly from the facts that the Northern gyre serves as a stabilizing mechanism (as just described) whereas a warmer South would decrease the density of the \textit{S} box (so increasing $q$) and $F_N$ is a destabilizing freshening into \textit{N}. The role of $\kappa$ is ambiguous, as a stronger feedback mechanism would stabilize the AMOC via the thermal circulation, and destabilize it via the haline circulation (the salt-advection feedback).

The AMOC collapse paradoxically starts off with an increase in its strength. Initially the optimal hosing scheme salinifies the northern \textit{N} and the thermocline \textit{T} boxes, while the other salinities change marginally as their volumes are larger. The increase in $S_N$ causes an increase in $q$, peaking at roughly $t \approx 10.8$ years with $q \approx 16.3$Sv. This short salinification pulse ends already at $t \approx 12$ years, and $q$ is back at its original level. However, the current state is a transient state (i.e. non-equilibrium), with box $T$ slightly fresher ($\mathcal{O}(10^{-3})$), while the others are marginally saltier ($\mathcal{O}(10^{-4})$) compared to the equilibrium ON-state. So the salt is redistributed. The salinity difference $S_N-S_S$, which determines the AMOC strength, is still the same, while $S_T-S_N$ has decreased. 
When applying freshwater to the Northern boxes the salinity difference $S_N-S_S$ drops, and $S_N$ cannot as easily be restored via the gyre since $S_T-S_N$ is lower. This also follows from \protect\eqref{eq:stabroot}: as $\phi_2-\phi_3$ decreases, the discriminant $D_{\text{ON}}$ will decrease, thus the distance between the two roots in Fig.~\protect\ref{fig:parabola}a diminishes and the stability region for an ON-state narrows. This destabilizing effect of the initial AMOC strengthening is quantified by computing the probabilities of a collapse within 100 years for trajectories starting in the standard ON-state and in this destabilized state. These probabilities are estimated using Monte Carlo simulations for various (unrealistically high) noise levels $\sqrt{\nu} \in \{0.11, 0.10, 0.09, 0.08\}$Sv. They are $\{(6.4\cdot10^{-3},6.5\cdot10^{-3}), (1.9\cdot10^{-3}, 2.1\cdot10^{-3}), (3\cdot10^{-4},4\cdot10^{-4}), (2\cdot10^{-5},4\cdot10^{-5})\}$ for trajectories starting in the ON-state and in the destabilized state respectively. Hence this probability is only slightly larger in the destabilized state than in the ON-state for the various noise levels. However, the ratio of the respective probabilities increases for smaller noise, and hence especially in the low-noise regime this initial strengthening of the AMOC increases the collapse probability significantly.

After this initial strengthening, a relatively large freshwater pulse is added to \textit{T} during $t \in (11,15)$ years causing its salinity to drop. Now fresher water is transported northward by the AMOC, and so $S_N$ drops too. This in turn causes the AMOC strength to decrease, so less saline water is transported northward by it, which decreases $S_N$ even further: the salt-advection feedback. This freshening of \textit{N} is partly offset by salt transport by the Northern gyre. Note that this also follows from \protect\eqref{eq:qdoton}: for increased $\phi_2$ the stable root disappears and $\dot{q}_{\text{ON}}<0$ for all $q>0$ (Fig.~\protect\ref{fig:parabola}b). Using \protect\eqref{eq:stabroot} and taking $\phi_3 \approx \phi_{\text{ON,}3}$ then no roots exist for $\phi_2 < 1.0146$ ($S_T < 35.5$). During this time the salinities of \textit{S} and \textit{IP} increase due to the pulse of salinity, and remain at this level with an increased Indo-Pacific gyre transport.

The freshwater pulse peaks at $t = 12$ years and then decreases. The \textit{T} box is not freshened anymore and its salinity increases again due to the reduced advection of saline water to the north and of fresher water from the \textit{S} and \textit{IP} boxes. Apparently a fresher Atlantic thermocline is no longer needed to destabilize the AMOC. Its salinity even overshoots before it converges to its new stable $\phi_{\text{OFF,}2}$ state. As can be seen from figure \protect\ref{fig:parabola}c a stable ON-state exists again, but the current state of the AMOC is to the left of the unstable root and hence in the unstable regime. The AMOC is now too weak to sustain itself as it cannot transport enough salt northward and even a saltier thermocline cannot save it. At the maximum value of $\phi_2 \approx 1.025$ ($S_T\approx 35.9$) the AMOC is still unstable ($\dot{q}<0$) for $q<2.9$Sv.

In Fig.~\protect\ref{fig:parabola}d the final phase of the collapse is depicted: a sharp freshwater pulse has put the AMOC into the unstable regime. It will continue to the OFF-regime ($q<0$) and eventually collapse to the stable OFF-equilibrium at $q=-6$Sv. Apparently it is more efficient to put the AMOC in this regime via a salinification and a subsequent freshwater pulse than via just one big freshwater pulse. Note that in this new regime ($q<0$) there is only one stable root  (Fig.~\protect\ref{fig:parabola}e), so from now on it converges deterministically to this stable OFF-state and the freshwater forcing is no longer needed. 

\subsubsection{Recovery}
\begin{figure*}
    \centering
    \includegraphics[width = 39pc]{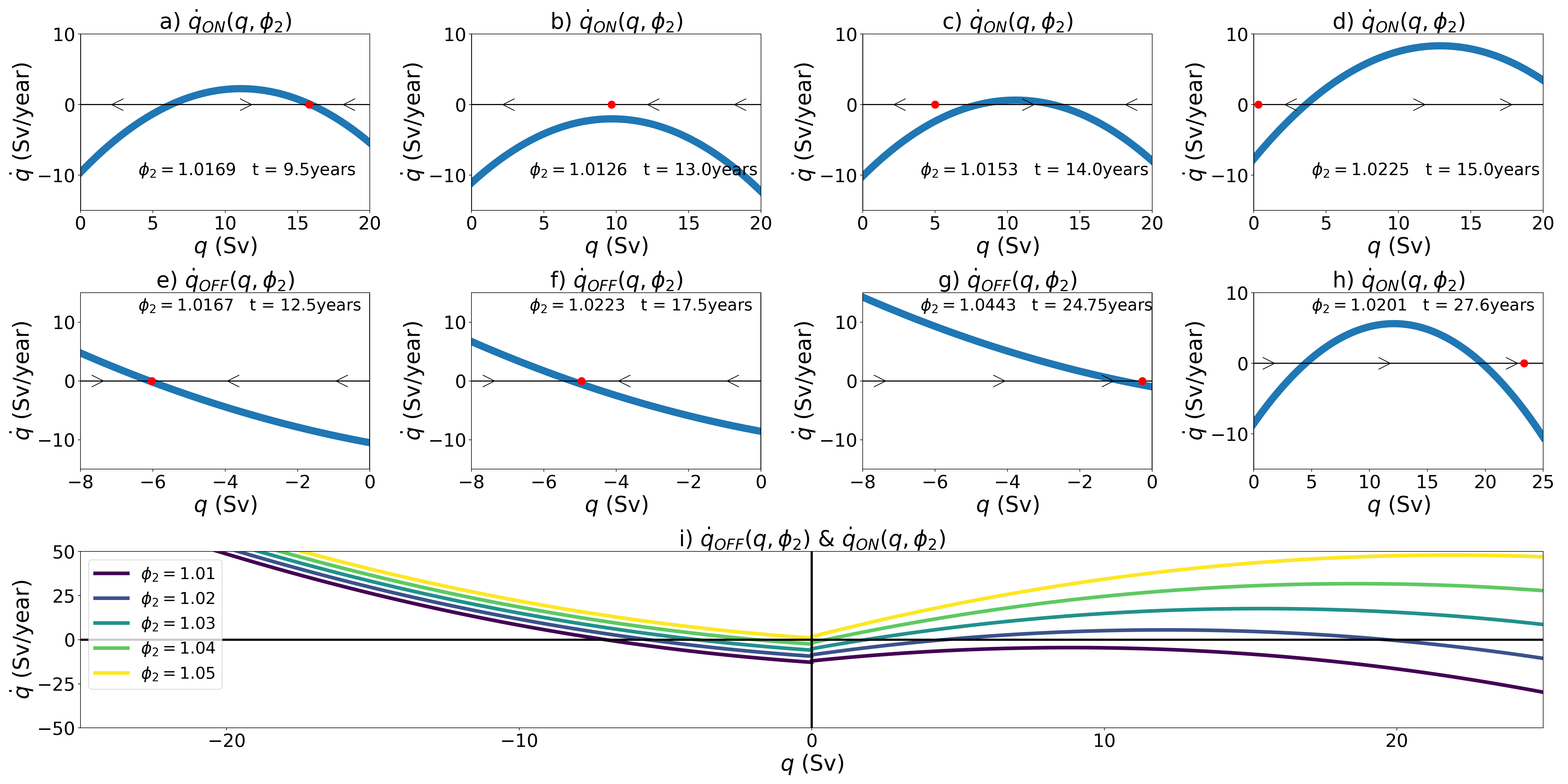}
    \caption{Dynamics of the AMOC collapse (recovery) in the top (middle) figures: $\dot{q}(q,\phi_2)$ (blue) in the $(\dot{q},q)$-plane with time $t$ and salinity $\phi_2$ values along each instanton added, and current AMOC strength $q$ indicated by the red dot. The arrows on the $q$-axis indicate the sign of $\dot{q}$. Bottom figure $i)$ shows complete $\dot{q}(q,\phi_2)$ curves for several $\phi_2\in(1.01,1.05)$. }
    \label{fig:parabola}
\end{figure*}

As seen in  figure \protect\ref{fig:parabola}e in the OFF-regime ($q<0$) there is only one root (equilibrium), which is stable. The physical intuition here is that for a lower (higher) AMOC, more (less) salinity is transported from \textit{B} to \textit{N}, which salinifies (freshens) \textit{N} and hence increases (decreases) the AMOC.

In order to restart the AMOC the \textit{N} box needs to be more saline, which can only be achieved via Northern gyre transport from \textit{T} and its increased salinity $S_T$. That is why $\phi_2$ only occurs multiplied by the gyre coefficient $K_N$ in \protect\eqref{eq:qdotoff}. From this equation it can also be derived that there will always be a stable attracting root in the range $q<0$ unless
\begin{equation}\label{eq:restartAMOC}
    \phi_2 \geq \phi_3 + \frac{f_N}{k_N}-r(T_S-T_0) \approx 1.045.
\end{equation}
For any higher $S_T$ values, the gyre transport can salinify the \textit{N} box enough for the AMOC to restart. From \protect\eqref{eq:restartAMOC} it also follows that an increased gyre coefficient $K_N$ or temperature difference would aid this recovery, whereas increasing $S_S$ or the freshwater flux $F_N$ would 
inhibit it.

So the salinities $S_T$ and $S_N$ (via the gyre transport) slowly increase until $S_T \approx 36.58$ ($\phi_2\approx 1.045$). Meanwhile in the other boxes the salinity remains largely unchanged. As $S_N-S_S$ slowly increases, so does the AMOC strength, see Figs.~\protect\ref{fig:parabola}f-g. Note that the AMOC remains close to the shifting stable root (the equilibrium).

The moment that the AMOC reverses, indeed at $S_T \approx 36.58$, there is no longer a attracting stable OFF-state. Suddenly a large salinity transport occurs from the \textit{T} to the \textit{N} box. Hence the salinity in the \textit{T} box rapidly drops, while the opposite takes place in the Northern box. With this strong increase in $S_N$ the AMOC strength also rises fast, and even overshoots above the level of the stable root $\dot{q}_{\text{ON}}(q,\phi_2) = 0$ (Fig.~\protect\ref{fig:parabola}h). During the increase of the AMOC,  the salinities in  \textit{IP} and \textit{S} adjust with a decreased Indo-Pacific gyre transport. 

From thereon the AMOC  converges to the stable ON-state: the \textit{N} box freshens and  the \textit{T} box salinifies via a reduced Northern gyre transport, with that the AMOC strength decreases to the stable ON-equilibrium at $q = 15$Sv (Fig.~\protect\ref{fig:overview_every}d). Note that due to high salinity $S_T$ just after $q>0$ is attained, the basin of attraction of the stable root is huge, and so the system can reach this equilibrium without any additional forcing (Fig.~\protect\ref{fig:parabola}i). 

\subsection{Sharp vs Sustained Forcing}
There is a stark asymmetry between the collapse and the recovery of the AMOC. For the collapse a sharp pulse of freshwater is sufficient to get the system into the unstable regime from whereon it collapses due to its own dynamics. So the pulse already decreases before the actual collapse at $q=0$. In the OFF-region there is no unstable regime, and the stable root has to be shifted with a sustained pulse of salinity until it is close to the AMOC restart at $q=0$.

The differences between equations \protect\eqref{eq:qdoton} and \protect\eqref{eq:qdotoff} are between the coefficients $B_{\text{ON}}$ and $B_{\text{OFF}}$, and that $A_{\text{ON}} = -A_{\text{OFF}}$. The latter equation follows from the fact that for a too large AMOC ($|q|\to\infty$) it transports a lot of salinity out off \textit{N} box to either \textit{B} or \textit{T}, and so $S_N$ decreases again (note that $\dot{\phi}_1\sim-|q|\phi_1$). Hence $\dot{|q|}\sim -|q|^2$ prevents $|q|$ from exploding. The difference between the coefficients $B_{\text{ON}}$ and $B_{\text{OFF}}$ is more interesting as the thermocline's salinity appears in $B_{\text{ON}}$ but not in $B_{\text{OFF}}$. This is because in the ON-regime the salinity $S_N$ (and so the AMOC strength) is also determined by the advective salinity transport from the thermocline (i.e. $q\phi_2$). In the OFF-regime only the gyre transport ($K_N\phi_2$) is present. Hence, in the ON-regime it is possible to use $\phi_2$ to manipulate the AMOC  to enlarge the unstable regime, whereas for the OFF-regime the $B_{\text{OFF}}$ term cannot be altered.

Lastly, we consider the shape of the optimal forcings associated with the most likely transition paths. A sharp pulse is optimal to collapse the AMOC, and to recover a gradual (almost linear) increase in forcing has the highest probability. In both cases a certain salinity value $S_T$ needs to be attained: for the collapse \textit{T} needs to be fresh enough such that $\dot{q}_{\text{ON}}$ has no roots, while for the recovery it needs to be saline enough to move the stable root to $q=0$. Also for both cases there is a time penalty: the longer it takes to get to this $S_T$ value the more of the input leaks away to other boxes. On the other hand, it should not be done too fast, because this requires high freshwater input values, which are unlikely. The additional leakages that occur when changing $\phi_2$ are:
\begin{equation*}
    V_2\frac{V_0}{t_d}\derp{\dot{\phi}_2}{\phi_2} = -(K_N+K_S)-|q|
\end{equation*}
which are additional outpour via gyres, and varying outpour via advection. Now, $K_N+K_S\approx 10.9$Sv, and $|q|\approx15.5$Sv and $|q|\approx6.3$Sv in the stable ON and OFF-state respectively. So in the recovery, the additional outpour from the thermocline box is mainly via the gyres, especially as $|q|\to0$. Therefore the additional salinity input during the recovery ideally grows linearly, since the leakage also grows linearly. However, for the collapse the decreased salinity transport out off the thermocline is dominated by the AMOC advection. As it diminishes, less salinity is transported out off the Atlantic thermocline and it becomes more saline, i.e. as $|q|\to0$ the salinification rate $\dot{\phi}_2$ slows down with the freshening of box \textit{T}. The evolution of the AMOC is non-linear due to salt-advection feedback and it can be heuristically concluded (as the shape of the optimal freshwater forcing cannot be analytically determined) that the loss of advective salinity transport is compensated by the surface freshwater input, and hence the initial shape of the optimal freshwater pulse is determined by the decrease of the AMOC.

\section{Response to Surface Freshwater Forcing} \label{sec:hoos}

\subsection{Effect of Freshwater Forcing Locations}
Following the Large Deviation Principle (LDP) the leading order term of the ratios of probabilities 
in the limit of small noise can be determined. Suppose one has paths $\phi_A(t)$ and $\phi_B(t)$ with Freidlin-Wentzell actions $S_A$ and $S_B$ respectively then their relative likelihood is
\begin{align*}
    \frac{\Prb(\phi_B)}{\Prb(\phi_A)} 
    \asymp\exp[(S_A-S_B)/\epsilon], 
\end{align*}
as long as $\phi_A$ and $\phi_B$ occur in systems like \protect\eqref{eq:genSDE} with the same dimensions $d$ and $k$ and the same smallness parameter $\epsilon$, see the derivation in Appendix C. Now the relative likelihood for the collapse and recovery  of the AMOC can be determined: $S[\phi_{\text{ON}\to\text{OFF}}] = \frac{1}{2}\sum_t\xi^2_{\text{ON}\to\text{OFF}}\Delta t = 0.00865$ and $S[\phi_{\text{OFF}\to\text{ON}}] 
= 0.01131$ for the standard case of FAMOUS\_B calibrated parameters. Therefore the collapse is $\exp[0.00266/\epsilon]$ times more likely than its recovery, which is a factor 
$1.645\pm0.115$ for our value of $\epsilon = 0.0055\pm0.0008$. 

\begin{table*}
\centering
\begin{tabular}{@{}l@{}lccccc}\toprule
 &&\multicolumn{5}{c}{$\gamma_2$} \\
 \cmidrule{3-7}
     && 0 & 0.25 & 0.5 & 0.75 & 1\\
 \cmidrule(lr){3-3}\cmidrule(lr){4-4}\cmidrule(lr){5-5}\cmidrule(lr){6-6}\cmidrule(lr){7-7}\multirow{5}{1cm}{$\gamma_1$}
 &0&$0.71\pm0.04$&$0.73\pm0.04$&$0.62\pm0.04$&$0.44\pm0.05$&$0.27\pm0.05$\\
 &0.25&\textbf{1.41}$\pm$\textbf{0.07}&\textbf{1.45}$\pm$\textbf{0.08}&\textbf{1.30}$\pm$\textbf{0.05}&\textbf{1.03}$\pm$\textbf{0.01}&$0.72\pm0.03$\\
&0.5&\textbf{1.10}$\pm$\textbf{0.02}&\textbf{1.13}$\pm$\textbf{0.02}&\textbf{1.06}$\pm$\textbf{0.01}&$0.91\pm0.02$&$0.72\pm0.04$\\
 &0.75&$0.64\pm0.05$&$0.66\pm0.04$&$0.65\pm0.04$&$0.58\pm0.05$&$0.51\pm0.05$\\
 &1&$0.37\pm0.05$&$0.38\pm0.06$&$0.37\pm0.05$&$0.37\pm0.06$&$0.33\pm0.05$\\
\bottomrule
\end{tabular}
\caption{The ratio of probabilities of an AMOC collapse under various hosing parameters $(\gamma_1,\gamma_2)$ compared to the collapse under standard FAMOUS\_B calibrated hosing parameters for $\epsilon = 0.0055\pm0.0008$. The fractions of the hosing are $(A_N,A_T,A_S,A_{IP}) = (\gamma_1, 1-\gamma_1, -\gamma_2, \gamma_2-1)$, so increasing $\gamma_1$ results in more freshwater hosing in the \textit{N} box and less into the \textit{T} box, and increasing $\gamma_2$ results in more salinity compensation in the \textit{S} box and less into the \textit{IP} box. The bold values indicate an increased likelihood of an AMOC collapse.}
\label{tab: probrat}
\end{table*}

A similar analysis can be done for the AMOC collapse under various freshwater flux parameters as a way to quantify the effect of different forcing strategies. In \protect\citet{smith2009study} it is shown that the AMOC response is sensitive to the choice of hosing regions. By varying the fraction of freshwater forcing in different surface boxes, we can express this sensitivity in terms of relative likelihoods. Previously we used the noise parameters from the FAMOUS\_B calibration, but now we vary these and compute for each different set of $\big(A_N, A_T, A_S, A_{IP}\big)$ the most likely noise-induced collapse and associated noise. These parameters are changed using 
\begin{equation*}
    \big(A_N, A_T, A_S, A_{IP}\big) = \big(\gamma_1, 1-\gamma_1, -\gamma_2, \gamma_2-1\big)
\end{equation*} 
with varying $0\leq\gamma_1,\gamma_2\leq1$. Next, the noise vector $\sigma$ is computed as 
\begin{equation*}
    \sigma_i = f_{\text{norm}}\frac{A_i}{V_i}
\end{equation*}
with normalization factor 
\begin{equation*}
    f_{\text{norm}} = 0.1063181\Big/\sqrt{\sum_i\big(\frac{A_i}{V_i}}\big)^2
\end{equation*} such that the norm $\|\sigma\|$ is $ 0.1063181$ as is the case for the original parameters, and a fair comparison can be made. Following these computations, the earlier used $\big(A_N, A_T, A_S, A_{IP}\big) = \big(0.070, 0.752, -0.257, -0.565\big)$ is achieved using $(\gamma_1,\gamma_2) = (0.09, 0.31)$. The results are in Table~\protect\ref{tab: probrat}, where the ratio of the probability of an AMOC collapse for various $(\gamma_1, \gamma_2)$ values compared to the probability of an AMOC collapse under the standard parameters are listed.  


From Table~\protect\ref{tab: probrat} it follows that to increase the likelihood of an AMOC collapse the freshwater forcing needs to be redistributed with 
a higher fraction of fresh water into the northern box (increased $\gamma_1$) and a lower fraction of compensation into the southern box (decreased $\gamma_2$). 
However, these changes should not be overdone, as dumping all fresh water into the northern box, or compensating 
for it fully from the southern box would actually reduce the likelihood.

\begin{figure*}
    \centering
    \includegraphics[width = 39pc]{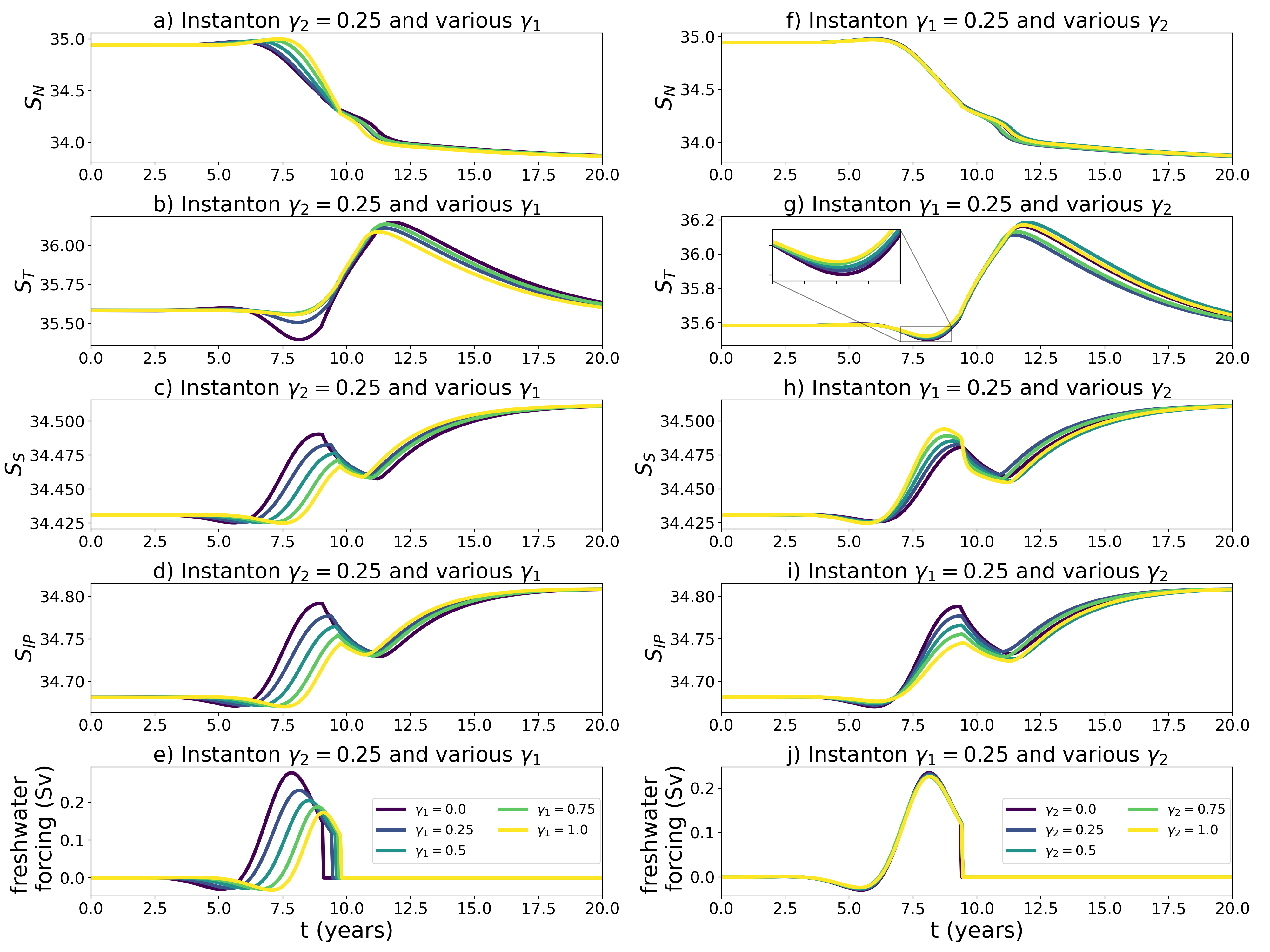}
    \caption{The instantons for varying $\gamma_1$(left) and varying $\gamma_2$(right) with dimensional salinities $S_N$ (a, f), $S_T$ (b, g), $S_S$ (c, h) and $S_{IP}$ (d, i) and the associated optimal freshwater forcings (e, j).}
    \label{fig:hosingvary}
\end{figure*}

To examine the physical reasons behind the changes in likelihood, the scenario with the highest likelihood of collapse ($(\gamma_1,\gamma_2) = (0.25,0.25)$), and scenarios with either one of the parameters changed are analyzed. Their 
instantons are shown in the Fig.~\protect\ref{fig:hosingvary}. 
Varying $\gamma_1$ yields the instantons in the left panels of Fig.~\protect\ref{fig:hosingvary}. The differences in the salinities $S_S$ and $S_{IP}$ between scenarios are marginal ($\mathcal{O}(10^{-2})$). As $\gamma_1$ increases a higher fraction of the freshwater is dumped into box \textit{N}, and so the AMOC $q$ shuts down more rapidly. The surplus salinity left behind in \textit{T} cannot leave via box \textit{N} and is then redistributed via gyres to $S$ and $IP$ and so the salinities here are slightly higher during the collapse. Moreover for lower $\gamma_1$ the \textit{T} box is fresher, since with higher $A_T$ more freshwater is simply deposited here. Similarly, as $\gamma_1$ increases the salinity $S_N$ drops more rapidly as a greater amount of freshwater is put here.

As $\gamma_1$ is higher than the optimal value $\gamma_1 =0.25$ the initial salinification becomes larger, while the subsequent freshwater pulse is sharper and higher. As discussed before the initial salinification puts the system into a transient state with a fresher \textit{T} box, which destabilizes the AMOC, see also derivation \protect\eqref{eq:qdoton}. As $\gamma_1$ increases, it becomes harder to directly freshen \textit{T}, and so the higher initial salinification with a temporary stronger AMOC is required. However, $S_T$ is still higher than usual, so the upcoming freshwater pulse also needs to be stronger. On the other hand, if $\gamma_1$ decreases from the optimal value the only way to get the necessary fresh water into box \textit{N} is via advection and gyre transport, which is ineffective as the freshwater input into \textit{T} also leaks to the \textit{S} and \textit{IP} boxes. So some freshwater input directly into \textit{N} helps, especially since $V_N$ is relatively small so its salinity is more easily changed. Concluding, for $\gamma_1$ there is an optimal compromise: $\gamma_1$ needs to be low enough to directly freshen the stabilizing saline thermocline, yet high enough to add freshwater to the relatively small \textit{N} box.

In the right panels of Fig.~\protect\ref{fig:hosingvary} again the optimal hosing scenario is taken, but now $\gamma_2$ is varied. 
Compared to varying $\gamma_1$ the salinity and likelihood changes across hosing scenarios are quite small. The salinity 
changes in the directly affected boxes \textit{S} and \textit{IP} are only of order $\mathcal{O}(10^{-2})$. The effects are straightforward: an increased $\gamma_2$ implies a higher fraction of salinification into box \textit{IP} and so $S_{IP}$ rises while $S_{S}$ drops. For the boxes \textit{N} and \textit{T} there are hardly any differences. Also the optimal freshwater perturbations change only marginally. 

The changes in likelihood can be derived from the graphs in Fig.~\protect\ref{fig:hosingvary}. As $\gamma_2$ is put to zero ($A_S = 0$) there is no way to directly control $S_S$ and so any required marginal changes are done via ineffective gyre and AMOC transport. So the likelihood slightly decreases. On the other hand, decreasing $A_{IP}$ has a significant negative effect on the likelihood of collapse. This can be understood with the zoomed inset in the $S_T$-graph (Fig.~\protect\ref{fig:hosingvary}g). As $A_{IP}$ decreases the southern box becomes more saline, while the Indo-Pacific freshens. The transport from \textit{S} to \textit{T} is larger than the one from \textit{IP} to \textit{T}, since $K_N + \gamma q \approx 11$ Sv while $(1-\gamma) q \approx 9$ Sv (taking $q\approx 15$ Sv). So a saltier Southern Ocean means a saltier Atlantic thermocline which stabilizes the AMOC (see \protect\eqref{eq:qdoton}) and hence decreases the likelihood of collapse. 

\subsection{Linear Deterministic versus Optimal Stochastic Forcing}
\begin{figure}
    \centering
    \includegraphics[width = 19pc]{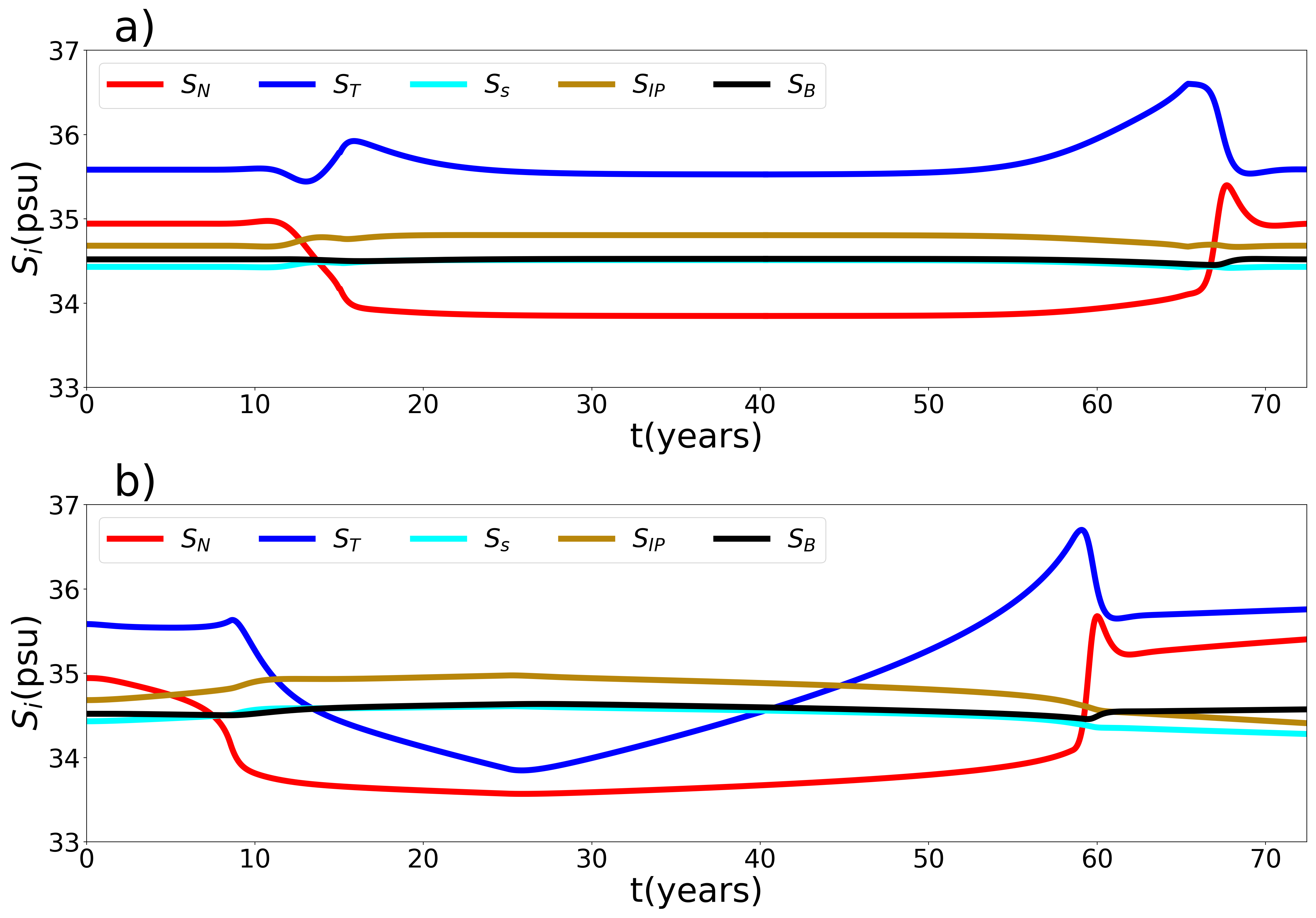}
    \caption{(Top, a) The instanton  directly followed by the `anti'-instanton, and (Bottom, b) the deterministic model behaviour under the prescribed freshwater forcing (\protect\ref{e:hosing_f})  for dimensional salinities $S_N$ (red), $S_T$ (blue), $S_S$ (cyan), $S_{IP}$ (brown) and $S_B$ (black).}
    \label{fig:hosing}
\end{figure}

In this section, the behaviour of the AMOC in the noise-induced transition following the instanton is compared to its deterministic behaviour in a bifurcation-induced tipping. For the latter, we force the AMOC with a prescribed freshwater flux which is piecewise linear time dependent similar to the method of \protect\citet{wood2019observable}.  
The goal there was to force the AMOC to stay close to equilibrium and hence the bifurcation 
tipping was produced. In Fig.~\protect\ref{fig:hosing}b,  the graph from \protect\cite{wood2019observable} 
is qualitatively reproduced with the following deterministic hosing procedure:
\begin{equation*}
    \dot{\phi}(t) = f(\phi(t)) + \sigma h(t)
\end{equation*}
where
\begin{align}
    h(t) = 
    \begin{cases}
        &-0.012 t \quad\quad\hspace{1.32cm}\text{ for }t<25\\
        &-0.3 + 0.012(t-25)\quad\text{ for }t>25.
    \end{cases}
\label{e:hosing_f}
\end{align}
where $\sigma$ follows from the original hosing parameters from the FAMOUS\_B calibration. Note that with this simple hosing procedure, the amount of hosing sufficient for a transition is not minimized nor the stable OFF-state was reached. Moreover, the graph in the original paper was produced with the FAMOUS\_A hosing parameters.  
In Fig.~\protect\ref{fig:hosing}a the instanton is shown, directly followed by the `anti'-instanton, 
so it is the most likely transition path from ON to OFF and back. Both are computed for the original hosing fractions and already discussed in section 4. The comparison of the two graphs in Fig.~\protect\ref{fig:hosing} illuminates the different behaviors of 
a noise-induced and a bifurcation-induced tipping.

Firstly, the salinity $S_T$ is in the deterministic hosing simulation much lower than for the instanton trajectory. In the former fresh water is continuously put into the thermocline box causing its salinity to stay low. But along the instanton one sees $S_T$ bouncing back, since an AMOC collapse means that salt is no longer transported northward out off the thermocline.
Secondly,  the salinity $S_N$ drops less rapidly than in the instanton's case. The main reason is that in the latter case a sharp fresh water pulse is introduced to the \textit{T} box, which rapidly freshens the northern box via the AMOC before it collapses, whereas in the former case the fresh water is introduced gradually so $S_N$ drops less rapidly.
Thirdly, smaller features  are seen in the trajectories of $S_{IP}$ and $S_S$. For these boxes the salinities grow more slowly since fresh water is extracted more gently. It seems that due to the excess fresh water extraction in the hosing the salinities of the these two boxes and of the \textit{B} box is artificially higher.
Lastly, the recovery of the AMOC shows the same overshooting for both freshwater forcing strategies: in both cases we need $S_T \approx 36.58$, since this is just salty enough to reactivate the AMOC. After this, all the salt is suddenly transported northward, causing the huge rise and drop in $S_N$ and $S_T$ respectively. Small salinity differences after the recovery between the hosing strategies are because in the linear hosing case fresh water is still being extracted from \textit{T}, and so additional salt is transported northward causing an higher $S_N$. That the bifurcation and noise-induced tipping do not differ that much when the AMOC recovers, could also be concluded from figures \protect\ref{fig:parabola}e, f and g, as the system stays close to the shifting stable root. 

Comparing the bifurcation points of the system with the computed optimal forcings also indicate the similarity between the most likely noise-induced and bifurcation-induced recovery. The AMOC has two stable states for $h\in(-0.069, 0.025)$, i.e. $(-0.066,0.180)$ Sv of freshwater input into \textit{N} and \textit{T}. So in the quasi-equilibrium approach done in Fig.~\protect\ref{fig:hosing}b the AMOC is in a mono-stable OFF-regime and ON-regime for $t\in(6,44)$ years and $t>52$ years respectively. The optimal forcing for the noise-induced AMOC collapse only crosses the bifurcation point briefly for $1.7$ years ($t\in(12,13.7)$ years, see Fig.~\protect\ref{fig:overviewONOFF}c) and this ends already $1.3$ years before reaching the separatrix. A brief time interval where no unstable ON-state exists (see Fig.~\protect\ref{fig:parabola}b) is enough to put the AMOC in an unstable regime from whereon it collapses even though a stable ON-state exists again (see Fig.~\protect\ref{fig:parabola}c). Contrarily, the optimal forcing for the recovery crosses the bifurcation threshold for $t\in(15.9, 25)$ years continuing till the separatrix, see Fig.~\protect\ref{fig:overviewONOFF}d. So here the separatrix is crossed when no stable OFF-state exists, similar to a bifurcation-induced tipping. This can also be deduced from Fig.~\protect\ref{fig:parabola}e, f and g: the crossing happens when eventually there is no stable root for $q<0$ i.e. the OFF-regime no longer sees any state with a decreasing $q$ for the given salinity $\phi_2$. Hence, the most likely noise-induced collapse only crosses the bifurcation briefly and before reaching the separatrix, while the noise-induced collapse has a sustained forcing that crosses this bifurcation threshold through to the separatrix and behaves like a bifurcation-induced tipping.

\section{Summary and Discussion} \label{sec:con}
In this work we computed the most likely paths of an AMOC collapse and recovery in a stochastic version of the \protect\citet{wood2019observable} model. This a conceptual box-model representation of the World Ocean which is calibrated to the AOGCM FAMOUS. We added the stochasticity to allow for noise-induced tipping of the AMOC, which was justified as a representation of the $P-E$ annual variation in the North Atlantic. Thanks to the model's simplicity the instantons can be readily computed. Using
this method we made a thorough analysis of noise-induced transitions of the AMOC. The most likely collapse of the AMOC will be preempted by an initial strengthening of the AMOC, which puts it into a less stable transient state. This is followed by a sharp freshwater pulse into the Atlantic thermocline such that the AMOC ends up in an unstable regime from whereon it collapses due to its own deterministic dynamics. Contrarily, the noise-induced recovery is quite close to the shifting equilibrium and is comparable to a bifurcation tipping. 

By comparing the actions of the instantons the ratios of their probabilities in the low-noise limit were computed. Any biases towards an ON or OFF-state due to the hosing method are now eliminated, accommodating a quantitative analysis of various hosing methods. This is a strength of the method since probabilities are determined and not merely mechanisms or theoretical possibilities of the hosing methods. This showed that to maximize the probability of an AMOC collapse most of the freshwater forcing needs to be added to the Atlantic thermocline, such that it cannot salinify the northern Atlantic, while only a small fraction of the compensation needs to be in the Southern Ocean since from here the Atlantic thermocline can most easily be salinified. 

The limitations of our approach are firstly the assumptions on the freshwater noise. Adding a white noise process for freshwater forcing to the model has been done many times \protect\citep{cessi1994simple, timmermann2000noise, castellana2019transition} and simplifies many computations, but under climate change this noise can be coloured and state-dependent. Furthermore, the system needs to be in the limit of low noise. This is the case for the model used here, and arguably holds for other AMOC models, as the noise compared to the AMOC strength should not depend too much on the model choice. The computation itself of the instanton can be tedious for more detailed ocean models and was here limited to low-dimensional continuous models. However, formally the Freidlin-Wentzell action holds for infinite dimensional stochastic PDE's  (SPDEs), and can be extended to multiplicative and even non-Gaussian noise \protect\citep{grafke2019numerical}. 

Lastly, this study's relevance relies on the present-day AMOC being in a multi-stable regime in order for noise-induced transitions to be possible. As mentioned it is debatable whether the observed negative AMOC induced freshwater transport at the southern boundary of the Atlantic basin is an indicator of this multi-stability. We think that the evidence in favour of this claim is strong enough that the possibility of multi-stability should not be ignored. A range of studies have shown this indicator to be reliable in a hierarchy of ocean models \protect\citep{de2005atlantic, dijkstra2007characterization, huisman2010indicator} and even in a state-of-the-art global climate model \protect\citep{van2023asymmetry}. However, there are also numerous studies arguing the present-day AMOC to be mono-stable \protect\citep{jackson2013shutdown, jackson2022understanding}. That said, we should consider that this indicator in these models is sensitive to model biases \protect\citep{mecking2017effect}. This debate is far from settled and outside the study's scope.

It will obviously be a  challenge to compute instantons in spatially extended models 
of the AMOC that are governed by SPDEs
, which would provide the identification of the most vulnerable freshwater forcing regions to tip 
the AMOC. From this, likely ensemble simulations can be designed to optimize the probability of a noise induced  
AMOC transition, which can possibly be used in global climate models. Also  --given a computed probability of 
an AMOC collapse \protect\citep{castellana2019transition}--  the probabilities for other forcing scenarios in the low-noise limit can 
be computed. Lastly, as instantons have been used as precursors for noise-induced 
tipping in more simplified models \protect\citep{giorgini2020precursors}, it is worth exploring whether they can be 
used to identify early warning signals of an approaching AMOC collapse. 

\appendix[A]
\appendixtitle{Numerical solution of the instanton equations}
 We want to solve the Hamiltonian equations \protect\eqref{e:Instantoneq} with boundary conditions $\phi(0) = x$ and 
 $\phi(T) = y$. Reformulating it as a minimization problem of the cost function yields:
\begin{align*}
    J\big(\phi(t),\theta(t),\mu(t)\big) &= \frac{1}{2}\int_0^T\langle \theta(t), a\theta(t)\rangle\,dt\\
    &+\int_0^T\langle\mu(t),\dot{\phi}(t)-f(\phi(t))-a\theta(t)\rangle\,dt\\
    &+\langle\beta,\phi(T)-y\rangle + \lambda\|\phi(T)-y\|^2
\end{align*}
where $\mu:\,[0,T]\to\R^d$ and $\beta\in\R^d$ are Lagrange multipliers to fulfill the constraints and $\lambda$ is an additional penalty term. Equalizing the variational derivatives to zero:
\begin{align*}
    &\frac{\delta J}{\delta \theta} = 0 \iff a(\theta(t)-\mu(t)) = 0\\
    &\frac{\delta J}{\delta \phi} = 0 \iff -\dot{\mu}(t)-\big(\nabla f(\phi(t))\big)^T\mu(t) = 0\\
    &\frac{\delta J}{\delta \mu} = 0 \iff \dot{\phi}(t)-f(\phi(t)) - a\theta(t) = 0
\end{align*}
and for the constraints:
\begin{align*}
    \derp{J}{\lambda} &= \|\phi(T)-y\|^2 = 0\\
    \nabla_\beta J &= \phi(T) - y = 0.
\end{align*}
Minimizing this cost function together with initial condition $\phi(0) = x$ solves the instanton equations. This pseudo-code is used to find the minimum:
\begin{align*}
        &\textbf{Given initial }\phi^0(t),\,\theta^0(t),\,\mu^0(t)\\
        &\textbf{Repeat:}\\
        &\qquad\textbf{Repeat untill }\|\theta^k(t)-\mu^{k+1}(t)\|<\textbf{tol :}\\
        &\qquad\qquad \textbf{Integrate forward }\\
        &\qquad\qquad\quad\dot{\phi}^{k+1}(t) = f(\phi^k(t))+a\theta^k(t)\,\\
        &\hspace{2.1cm}\,\textbf{ with }\phi^{k+1}(0)=x\\
        &\qquad\qquad \textbf{Integrate backward }\\
        &\qquad\qquad\quad\dot{\mu}^{k+1}(t) = -\nabla f(\phi^{k+1}(t))^T\mu^k(t)\\
        &\hspace{2.1cm}\,\textbf{ with }\mu^{k+1}(T)=-(2\lambda(\phi^{k+1}(T)-y)+\beta)\\
        &\qquad\qquad\theta^{k+1}(t) = \theta^{k} - \alpha \big(\theta^k(t)-\mu^{k+1}(t)\big)\\
        &\hspace{2.1cm}\,\textbf{ with }\alpha = \max\big\{\alpha\big|J(\phi^{k+1},\theta^{k+1}(t),\mu^{k+1})\\
        &\hspace{2.6cm}<J(\phi^{k+1},\theta^{k},\mu^{k+1})\big\}\\
        &\qquad \beta \to \beta + \lambda(\phi^{k+1}(T)-y)\\
        &\qquad \lambda\to f_\lambda\lambda
\end{align*}
where $f_\lambda>1$ a factor to update $\lambda$. The algorithm searches for the end condition  $\theta(T) = -\beta$ such that $\phi(T) = y$. The equation for the path $\phi(t)$ is integrated forward in time, while that for the conjugate momentum is integrated backward in time to ensure numerical stability. To speed up convergence we added $\theta = 0$ after $q$ switches sign. This is justified as the model converges deterministically to the new stationary point after this switch following the analysis done in section \protect\ref{sec:trs}b.
\vspace{0.5cm}

\appendix[B]
\appendixtitle{Used Parameter Values}
In Table \protect\ref{tab:par} the parameter values of the non-dimensional  model are provided. The volumes and transports are written in units $10^6$m$^3$ and Sv respectively, such that most parameters have an order of magnitude $\mathcal{O}(10^{-2})-\mathcal{O}(10^{2})$. The parameters that do not fullfil this always occur in reasonable ratios: $t_d/V_0\sim 0.31536\,10^{-6}$m$^{-3}$s and $C/(S_0V_0)\sim127$.
\begin{table*}
    \centering
    \begin{tabular}{lll}
     \toprule
     \multicolumn{3}{c}{Parameter Values} \\
     \cmidrule{1-3}
     $V_0\,(10^6$m$^3)$   & $10^{10}$    & volume scaling parameter\\
     $V_1$   & $3.261$    & volume northern box\\
     $V_2$   & $7.777$    & volume Atlantic thermocline box\\
     $V_3$   & $8.897$    & volume southern box\\
     $V_4$   & $22.02$    & volume Indo-Pacific box\\
     $V_5$   & $86.49$    & volume bottom box\\
     $t_d\,(s)$   & $3.1536\cdot10^9$    & time scaling parameter\\
     $S_0\,$(psu)   & $35$    & salinity scaling parameter\\
     $C\,(10^6$m$^3$psu)   & $4.446304026\cdot10^{13}$    & total salinity in basin\\
     $\alpha\,$(kg/m$^3$K)   & $0.12$    & thermal coefficient\\
     $\beta\,$(kg/m$^3$)   & $0.79$    & saline coefficient\\
     $\gamma\,$   & $0.39$    & proportion Cold Water Path\\
     $\eta\,$(Sv)   & $74.492$    & S-B box mixing parameter\\
     $K_N\,$(Sv)   & $5.456$    & northern subtropical gyre coefficient\\
     $K_S\,$(Sv)   & $5.447$    & southern subtropical gyre coefficient\\
     $K_{IP}\,$(Sv)   & $96.817$    & Indo-Pacific gyre coefficient\\
     $F_N\,$(Sv)   & $0.384$    & northern freshwater flux\\
     $F_T\,$(Sv)   & $-0.723$    & Atlantic thermocline freshwater flux\\
     $F_S\,$(Sv)   & $1.078$    & southern freshwater flux\\
     $F_{IP}\,$(Sv)   & $-0.739$    & Indo-Pacific freshwater flux\\
     $T_S\,$(K)   & $4.773$    & southern box temperature\\
     $T_0\,$(K)   & $2.65$    & base temperature\\
     $\mu\,$($10^{-6}$m$^{-3}$sK)   & $0.055$    & heat transport coefficient\\
     $\lambda\,$($10^{6}$m$^{6}$/kg s)   & $27.9$    & MOC-density difference coefficient\\
     \hline
    \end{tabular}
    \caption{Used Parameter Values}
    \label{tab:par}
\end{table*}

\appendix[C]
\appendixtitle{Relative Likelihood in Low-Noise Limit}
Consider two stochastic systems as in \protect\eqref{eq:genSDE} with the same smallness-parameter $\epsilon$ indicating the noise amplitude:
\begin{align*}
    dX^1_t &= f_1(X^1_t)\,dt + \sigma_1\sqrt{\epsilon}dW_t^1\\
    dX^2_t &= f_2(X^2_t)\,dt + \sigma_2\sqrt{\epsilon}dW_t^2
\end{align*}
with $X_t^i\in\R^d$ the state vector, $f_i:\R^d\to\R^d$ the deterministic drift, $W_t^i$ a $k$-dimensional Wiener process, and $\sigma_i\in\R^{d\times k}$ distributing the noise for $i\in\{1,2\}$. We need $||\sigma_1|| = ||\sigma_2||$ in order for the noise amplitudes to be the same in both systems. Again we state equation \protect\eqref{eq:LDP} for both systems:
\begin{equation*}
    \epsilon\log\Bigg(\Prb\bigg[\sup_{t\in[0,T]}||X_t^i-\phi^i(t)||<\delta\bigg]\Bigg) = -S_T^i[\phi^i] + \mathcal{O}(\epsilon)
\end{equation*}
where $S_T^i$ is the Freidlin-Wentzell action for system $i$ and $\phi^i(t): [0,T]\to\R^d$ is a path in that system. The relative likelihood in the low-noise limit between path $\phi^1(t)$ in the first and $\phi^2(t)$ in the second system is now computed as
\begin{align*}
    \epsilon\log\Bigg(\frac{\Prb\big[\sup_{t\in[0,T]}\|X_t^1-\phi^1(t)\|<\delta\big]}{\Prb\big[\sup_{t\in[0,T]}\|X_t^2-\phi^2(t)\|<\delta\big]}\Bigg) = \\
    S_T^2[\phi^2] - S_T^1[\phi^1] + \mathcal{O}(\epsilon)
\end{align*}
and hence 
\begin{align*}
    \frac{\Prb\big[\sup_{t\in[0,T]}\|X_t^1-\phi^1(t)\|<\delta\big]}{\Prb\big[\sup_{t\in[0,T]}\|X_t^2-\phi^2(t)\|<\delta\big]}\asymp\\
    \exp\big[(S_T^2[\phi^2]-S_T^1[\phi^1])/\epsilon\big].
\end{align*}

%

\acknowledgments
We would like to thank Reyk B\"orner, Raphael R\"omer, Jason Frank and the two anonymous reviewers for their insightful suggestions. J.S. and H.A.D. are funded by the European Research Council through ERC-AdG project TAOC (project 101055096). T.G. acknowledges support from EPSRC projects EP/T011866/1 and
EP/V013319/1. For the purpose of open access, the authors have applied a
Creative Commons Attribution (CC BY) licence to any Author Accepted
Manuscript version arising from this submission.

%
%
\datastatement
The results can be readily reproduced using the described method and the stated parameter values.


\bibliographystyle{ametsocV6}
\bibliography{references}

\end{document}